# C$_3$N: a Two Dimensional Semiconductor Material with High stiffness, Superior Stability and Bending Poisson's Effect

*Haidi Wang,$^a$   Hong Wu,$^a$ and Jinlong Yang$^{*ab}$*

*$^a$Hefei National Laboratory of Physical Science at the Microscale, University of Science and Technology of China, Hefei, Anhui 230026, China*

*$^b$Synergetic Innovation Center of Quantum Information & Quantum Physics, University of Science and Technology of China, Hefei, Anhui 230026, China*

*Email: jlyang@ustc.edu.cn*

Recently, a new type of two-dimensional layered material, i.e. C$_3$N, has been fabricated by polymerization of 2,3-diaminophenazine and used to fabricate a field-effect transistor device with an on–off current ratio reaching 5.5 × 10$^{10}$ (Adv. Mater. 2017, 1605625). Here we have performed a comprehensive first-principles study mechanical and electronic properties of C$_3$N and related derivatives. Ab inito molecular dynamics simulation shows that C$_3$N monolayer can withstand high temperature up to 2000K. Besides high stability, C$_3$N is predicted to be a superior stiff material with high Young's modulus (1090.0 GPa), which is comparable or even higher than that of graphene (1057.7 GPa). By roll-up C$_3$N nanosheet into the corresponding nanotube, an out-of-plane bending deformation is also investigated. The calculation indicates C$_3$N nanosheet possesses a fascinating bending Poisson's effect, namely, bending induced lateral contraction. Further investigation shows that most of the corresponding nanotubes also present high Young's modulus and semiconducting properties. In addition, the electronic properties of few-layer C$_3$N nanosheet is also investigated. It is predicated that C$_3$N monolayer is an indirect semiconductor (1.09 eV) with strongly polar covalent bonds, while the multi-layered C$_3$N possesses metallic properties with AD-stacking. Due to high stability, suitable band gap and superior mechanical strength, the C$_3$N nanosheet will be an ideal candidate in high-strength nano-electronic device applications.

## Introduction

Since the first two-dimensional (2D) material, namely graphene, was successfully fabricated by Novoselov and Geim,[1] 2D materials research has rapidly risen to be one of the hot spots of condensed matter physics due to their festinating electronic, mechanical, optical or thermal properties[2–4]. For instance, many fantastic phenomena are discovered in graphene,[5] due to the presence of a Dirac-type band dispersion. However, the gapless feature, at the same time, also limits its application in electronic and optoelectronic devices. Subsequently, few layer black phosphorus[6–8] was successfully synthesized and predicted that it not only has a direct band gap of about 2.0 eV, but also has a high carrier mobility. Unfortunately, black phosphorene degrades readily when it exposes to the air.[9] At the same time, due to the small Young's modulus, the black phosphorene is too flexible to be applied in high mechanical environment.[10] Therefore, exploring new 2D semiconducting materials with high stability, strong mechanical strength and suitable band gaps are still a long-term target.

Recently, a hole-free 2D crystal consisted of carbon and nitrogen atoms, named C$_3$N, has been fabricated by polymerization of 2,3-diaminophenazine.[11] C$_3$N possesses an indirect band gap of 0.39 eV (PBE level) that can be tuned to cover the entire visible range by fabrication of quantum dots with different diameters. In addition, back-gated field-effect transistors made of monolayer C$_3$N display a high on/off ratio of 5.5×10$^5$. These results may make C$_3$N a very promising candidate material for future applications in electronics and optoelectronics at the nanoscale level. Therefore, a further examining of the mechanical and electronic structure properties will be helpful to further exploit other applications of this new type of 2D material.

In this work, we have conducted a comprehensive investigation on the mechanical and electronic properties of C$_3$N monolayer, the calculation suggests that C$_3$N is a stiff material with ultra-high Young's modulus, so do the corresponding nanotube. Due to the special geometric structure, a fascinating bending Poisson's ratio is found when an out-of-plane bending load is applied to C$_3$N nanosheet. In addition, the basic electronic structure properties of monolayer and multi-layer C$_3$N are also investigated, indicating that C$_3$N monolayer is an indirect semiconductor with strong polar covalent bonds, while the multi-layered C$_3$N possesses metallic properties with AD-stacking order.

## Computational methods

In this work, all the first-principles calculations are performed based on the Kohn-Sham density functional theory[12] (KS-DFT) as implemented in the Vienna ab initio simulation package[13] (VASP). The generalized gradient approximation within the Perdew-Burke-Ernzerhof[14] (PBE) functional form is used for the exchange-correlation energy. The plane wave basis sets with kinetic energy cutoff of 500 eV are used to expand the valence electron wave functions. For all structural relaxations, the convergence criterion for the energy in electronic SCF iterations and the Hellmann-Feynman force in ionic step iterations are set to be 1.0×10$^{-6}$ eV and 5.0×10$^{-3}$ eV/Å, respectively. In order to reduce the interaction between neighboring layers, a large vacuum space of at least 15Å is introduced along the z-axis. The Brillouin zone is represented by Monkhorst–Pack[15] special k-point mesh of 12×12×1 for geometry optimizations, while a larger grid (16×16×1) is used for SCF computations. Besides, HSE06 hybrid functional is used to obtain an accurate band gap.[16] van der Waals (vdW) correction proposed by Grimme (DFT-D2) is used due to its good description of long-range vdW interactions for multi-layered 2D materials[17–21]. As a comparison, the Becke88 van der Waals[22] (optB88-vdW) functional is also used for multi-layered structure. Ab inito molecular dynamics (AIMD) simulations with NVT ensemble are performed to assess the thermal stability of C$_3$N monolayer.

## Results and discussions

The optimized structure of C$_3$N monolayer is shown in Fig. 1a. The structure possesses P6/mmm symmetry (space group ID 191) with hexagonal lattice. The optimized lattice constants are a=b=4.862Å. The top view shows that the new phase is similar to graphene sheet, however, the six membered rings are composed by either C and N atoms or totally C atoms. Unlike blue phosphorene[23], silicene[24] and some other puckered materials[25,26], the C$_3$N has a flat structure from the side view. In this structure, all C and N atoms are sp$^2$ hybridized forming conjugated π bond. The C-C (1.40 Å) and C-N (1.40Å) bond lengths show pronounced characters of single bonds. The unit cell of C$_3$N contains 8 atoms as denoted by red box in Fig.1a in which the C to N ratio is 3:1.

Before studying the mechanical and electronic properties of C$_3$N, we should determine whether it is stable structure. To confirm the dynamical stability, the phonon dispersions of C$_3$N are calculated by using the finite displacement method as implemented in PHONOPY.[27] The calculated phonon dispersion curve shows no imaginary modes in the entire Brillouin zone, which confirms that C$_3$N is dynamically stable. Moreover, we also carried out ab initio molecular dynamics (AIMD) simulation with 4×4×1 supercell to judge the thermodynamic stability of C$_3$N monolayer. After heating at room temperature (1000K) for 5ps with a time step of 1ps, no structure reconstruction is found. Furthermore, we find the C$_3$N can also withstand temperature under 2000 K and melting point get close to 3000K, suggesting that C$_3$N monolayer has good thermal stability (See in Fig. S1 and S2 ESI).

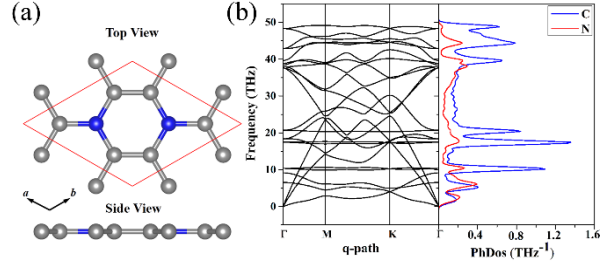

**Fig. 1.** (a) Top and side view of monolayer $C_3N$ in the unit cell. (b) Phonon band structure and partial DOS of $C_3N$.

Then we turn to discuss in-plane mechanical properties. Using the finite distortion method,[28] we calculate the linear elastic constants of $C_3N$. Due to the symmetry of geometric structure, there are three independent elastic constants for 2D hexagonal crystal, namely, $C_{11}$ ($C_{22}=C_{11}$), $C_{12}$ and $C_{44}$ (See Table 1). Therefore, the elastic constants of $C_3N$ satisfy the requirement of the Born criteria[29], namely $C_{11}>0$, $C_{66}>0$ and $C_{11}-C_{12}>0$, which further demonstrate the mechanical stability of $C_3N$. To have a deep knowledge of mechanical properties of $C_3N$, we plot the orientation dependent Young's modulus and the Poisson's ratio according to equation 1.[30]

$$\begin{cases} E(\theta) = \frac{Y_{zz}}{\cos^4\theta + d_2\cos^2\theta\sin^2\theta + d_3\sin^4\theta} \\ v(\theta) = \frac{v_{zz}\cos^4\theta - d_1\cos^2\theta\sin^2\theta + v_{zz}\sin^4\theta}{\cos^4\theta + d_2\cos^2\theta\sin^2\theta + d_3\sin^4\theta} \end{cases} \quad Equation\ 1$$

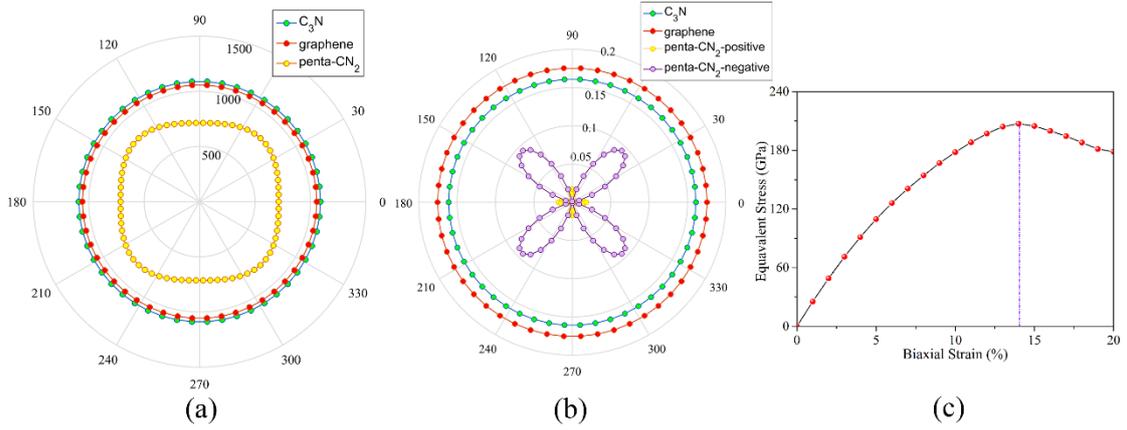

**Fig. 2.** Orientation dependent (a) Young's modulus (GPa) and (b) Poisson's ratio of $C_3N$ and graphene. (c) The strain-stress relation for monolayer $C_3N$. The strain is defined as $(a-a_0)/a_0$, where $a_0$ is the equilibrium lattice and $a$ stands for strained one.

As a comparison, the Young's modulus and the Poisson's ratio some representative 2D materials, such as graphene[31,32] and penta-$CN_2$[33], are also plotted Fig. 2a and 2b. Unlike penta-$CN_2$, the Young's modulus and the Poission's ratio of $C_3N$ is isotropic due to its high symmetry. More important than the isotropic mechanical properties, the Young's modulus of $C_3N$ (1090.0 GPa) is comparable or a little higher than that of graphene (1057.7 GPa) and much higher than that of penta-$CN_2$ (794.7 GPa), borophene (646.6 GPa)[34] and black phosphorene (166.0 GPa)[35], which suggests that 2D $C_3N$ is as stiff as graphene. To explore the ideally tensile strength (the highest achievable stress of a defect-free crystal at 0 K) and critical strain (the strain at which ideal strength reaches)[36] of $C_3N$, an in-plane biaxial tensile stress is applied. The stress-strain relationship for monolayer $C_3N$ is presented in Fig. 2c, where the tensile strain ranges from 0 to 20%. The ideal strengths under the critical strain (14%) is 207.1 GPa (Equivalent value 66.5 N/m), which is apparently larger than that of black phosphorene,[10] delta-phosphorene,[37] $MoS_2$[38] and borophene.[39] From the high Young's modulus and large ideal strength, we may conclude that $C_3N$ monolayer is a stiff material.

**Table 1.** Independent elastic constants ($C_{ij}$, GPa), Young's modulus (Y, GPa) and Poisson's ratio (v) of $C_3N$, graphene

and penta-$CN_2$ monolayer.

| System | $C_{11}$ | $C_{22}$ | $C_{12}$ | $C_{66}$ | Y | ν |
|---|---|---|---|---|---|---|
| $C_3N$ | 1119.0 | 1119.0 | 180.1 | 469.4 | 1090.0 | 0.16 |
| Graphene | 1091.3 | 1091.3 | 191.6 | 449.6 | 1057.7 | 0.18 |
| Penta-$CN_2$ | 714.0 | 714.0 | 12.9 | 438.4 | 713.7 to 794.7 | -0.09 to 0.02 |

To simulate the out-of-plane bending deformation of $C_3N$ monolayer, the $C_3N$ nanosheet is rolled into the corresponding nanotube, which has been widely studied in other 2D materials, such as graphene[40,41], boron nitride[42], blue phosphorene[43] and black phosphorene[44]. Here, monolayer $C_3N$ with flat structure is directly rolled up to $C_3N$-Nanotube. Similar to graphene, the structure information of nanotube is described by a pair of integer indexes (n,m) that defines a rollup vector **R**=n**a**+m**b** (see Fig. S3), where **a** and **b** are primitive lattice vector of $C_3N$ nanosheet. Specifically, two types of nanotube is considered in this work, namely, zigzag (n, 0) for n={3, 4,… 7} and armchair (-n, n) for n={2, 3,… 6}.

According to the Poisson's ratio formulates, lateral strains in a material can be caused by a uniaxial stress in the perpendicular direction, but no net lateral strain should be induced in a thin homogeneous elastic plate subjected to a pure bending load. Here, we find that significant exotic lateral strains can be induced while bending $C_3N$ sheet to corresponding nanotube (See Fig. 3a). Taking zigzag nanotube as an example, the lattice constant $r_0$ of all (n,0) tube should be equal to the width of $C_3N$ sheet without of bending load ($r_d$, Fig. 3a). However, as listed in Table 2, all of lattice constants $r_0$ are smaller than $r_d$=8.422 Å, indicating that $C_3N$ monolayer has an interesting bending Poisson effect[45]. To have deep knowledge of this behavior, the bending Poisson ratio ν, defined as the ratio of lateral strain to the curvature of $C_3N$-nanotube ($\nu=\frac{\varepsilon_L}{k}$) is evaluated, where the lateral strain introduced by the bending load can be written as $\varepsilon_L = \frac{r_d - r_0}{r_0}$ and corresponding lateral stress is defined as $\sigma_L = Y\varepsilon_L$. Y is the in-plane Young's modulus of $C_3N$ monolayer. According to the $\varepsilon_L - k$, $\sigma_L - k$ and $\nu - k$ relations illustrated in Fig. 3b-d, some typical features of the bending Poisson's effect in $C_3N$ monolayer can be summarized: I) The bending-induced lateral strain are orientation-dependent. For instance, when the radius of curvature down to ~2.6 Å, the lateral strain along the armchair direction is -0.45%, while the zigzag direction is up to 1.45%, which is easy to be detected by experiment. In addition, the zigzag nanotubes with small radius of curvature have a more obvious bending Poisson's effect than the armchair ones, however, with the increase of radius of curvature, the lateral strain of both armchair and zigzag nanotubes reduce to the 0. II) The variation trend of lateral stress under different radius of curvature is similar to that of lateral strain. Compared with in-plane strain-stress curve of $C_3N$ monolayer, the lateral stress is much small than in-plane strain-inducted equivalent stress. III) The bending Poisson ratio is a function of curvature. As for armchair nanotube, the $\nu - k$ relation is approximately linear, while zigzag nanotube has a nonlinear relation. To explain bending Poisson's effect, a local structure of $C_3N$ nanotube is shown in Fig. S4 (See in ESI). It can been seen that the perfect $C_3N$ sheet has a flat structure, where all carbon and nitrogen atoms hold $sp^2$ hybridization. However, when out-of-plane bending load is employed to the $C_3N$ monolayer, the $sp^2$ hybridization states are destroyed. To accommodate the bending load, the N atoms will prone to form $sp^3$ hybridization as it is in $NH_3$ molecule. Finally, the C atoms linked the same N atom will get close to each other and lead to an axial contraction.

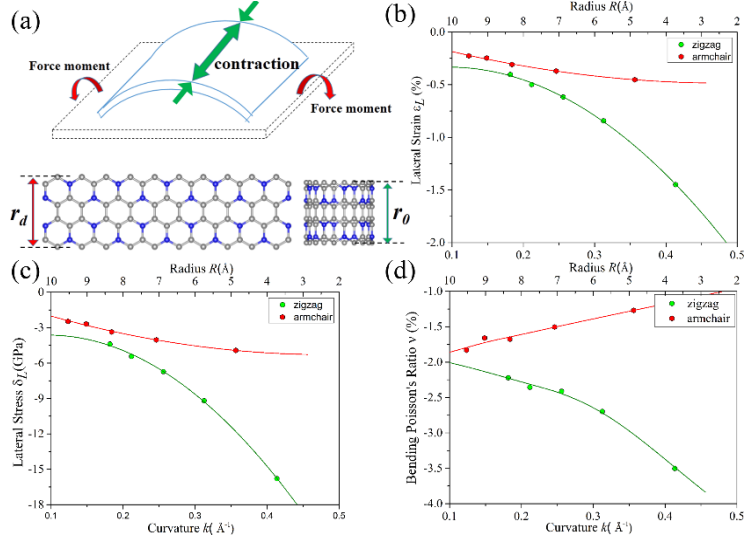

**Fig. 3. (a)** Schematic plot of homogeneous elastic plate under a pure bending loading. The plates with and without mechanical load are depicted by the blue and black dashed line. $C_3N$ nanosheet and corresponding nanotube are also included. **(b) (c)** and **(d)** Bending induced lateral strain, equivalent lateral stress and Bending Poisson's ratio as a function of curvature for bending along the zigzag (green line) and armchair (red line) directions.

Furthermore, the basic mechanical and electronic structure of $C_3N$ nanotube are also investigated (Table 2). For both armchair and zigzag $C_3N$ nanobube, as expected, the strain energy decreases with increasing $d$. The Young's modulus of all the nanotubes we selected are approximately equal to 1.0 TPa expcept the (3,0)-nanotube. It's worth noting that the Young's modulus of tubes are also comparable or a little higher than that of graphene. A qualltatively calculation shows that most of them are semiconductors with their bandgaps related with radius of curvature $d$. However, the zigzag nanotubes give rise to a much smaller band gap compared to that of the armchair nanotubes with nearly the same $d$. In detail, the band gap of (5,0)-nanotube is 0.25 eV, while that of (-3,3)-nanotube is up to 0.62 eV, which is even higher than that of flat $C_3N$ monolayer, indicating that bending load can be tool to tune he band gap of $C_3N$.

**Table 2.** The calculated diameter ($d$), lattice constant along the axial direction ($r_0$), bandgap under PBE level of theory ($E_{gp}$), Young's modulus ($Y$) and strain energy ($\delta E$). The capitalized letter M, I and D behind the band gap value stands for metal, indirect-semiconductor and direct-semiconductor, respectively.

| System | $d$(Å) | $r_0$(Å) | $E_{gp}$(eV) | $Y$(GPa) | $\delta E$(ev/atom) |
|---|---|---|---|---|---|
| (3,0) | 4.835 | 8.300 | 0.00/M | 958.3 | 0.236 |
| (4,0) | 6.398 | 8.351 | 0.01/I | 1010.9 | 0.138 |
| (5,0) | 7.814 | 8.370 | 0.25/D | 1047.4 | 0.087 |
| (6,0) | 9.444 | 8.380 | 0.35/D | 1045.8 | 0.060 |
| (7,0) | 11.006 | 8.388 | 0.41/D | 1045.2 | 0.043 |
| (-2,2) | 5.612 | 4.840 | 0.38/D | 1046.4 | 0.214 |
| (-3,3) | 8.123 | 4.844 | 0.62/D | 1075.5 | 0.088 |
| (-4,4) | 10.866 | 4.847 | 0.66/D | 1071.2 | 0.046 |
| (-5,5) | 13.446 | 4.850 | 0.64/D | 1081.4 | 0.028 |
| (-6,6) | 16.184 | 4.851 | 0.61/D | 1080.5 | 0.019 |

Finally, we also pay our attention to the electronic structure properties of $C_3N$. The electronic band structure as well as density of states (DOS) of $C_3N$ monolayer are calculated under the PBE level of theory. As shown in Fig. 4a, unlike gapless graphene[46], $C_3N$ is predicted to be indirect band gap semiconductor with a gap of 0.39 eV, as the valence band maximum (VBM) and the conduction band minimum (CBM) are located at the M point and the Γ point in the Brillouin zone, respectively. The calculated bandgap value is consistent with previous work.[11] It is well-known that DFT within PBE level of theory underestimates the bandgap of semiconductors.[47] Therefore, a more accurate hybrid functional HSE06[16] is employed to correct the bandgap. We verify that the band dispersion profiles remain the same (See in Fig. S5 ESI), but the band gap value is increased to 1.09 eV. From the projected band structure and density of states (DOS), one can see that the VBM is a hybrid state of C-$p_z$ and N-$p_z$, however, the CBM is mainly contributed by $p_z$ orbital of C atoms, which are consistent with their electronic charge density (See in Fig. S6 ESI). By calculating the electron localization function (ELF)[48,49], the C-C and C-N covalent bonds can be identified in $C_3N$ (Fig. 4c). In addition, according to our Bader's charge analysis,[50] each C atom transfers about 1.3$e$ to N atoms, which implies a significant polarizability of the C−N covalent bonds. Compared with previously proposed N chains encapsulated in carbon nanotubes[51] (0.4$e$) and the penta-$CN_2$[33] (1.2$e$), the nitrogen atoms in $C_3N$ receive much more charge, indicating a much stronger interaction between C and N atoms, which to some extent explains why $C_3N$ holds high Young's modulus.

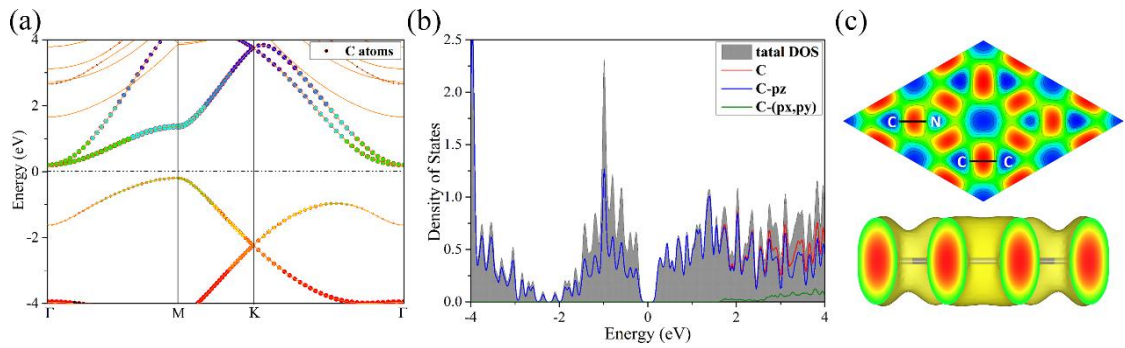

**Fig. 4.** (a) Electronic band structure and (b) projected density of state (PDOS). Γ(0.0, 0.0, 0.0), M (1/2, 0.0, 0.0) and K (-1/3, 2/3, 0.0) refer to special points in the first Brillouin zone. (c) The electron localisation function (ELF) of $C_3N$.

Finally, we also calculate the multi-layer $C_3N$ nanosheet. As for the geometric structure of bilayer $C_3N$, we here consider four kinds of high symmetry stacking order, namely AA-, AB-, AC- and AD-stacking. As shown in Fig. 5a, in AA-stacking, the top layer is directly stacked on the bottom layer and they are matched perfectly in xy-plane. The AB-, AC and AD-stacking can be viewed as shifting the top layer of atoms along the vector ***a-b*** with a displacement as displayed in Fig. 5b-d. As a comparison, the relative energy, layer distance and bandgap are listed in Table 3. We firstly optimize the structure by including DFT-D2 correction due to its good description of interlayer interaction. The calculated layer distance is in the range of 3.19-3.40 Å, which is close to the value of its parent material graphene[52] (3.35 Å) and analogue $C_2N$-h2D (3.18 Å).[53] By analyzing the relative energy, we may find that AD-stacking is the most favorable configuration for bilayer $C_3N$, being 0.093, 0.03 and 0.017 eV per unit cell lower than that of AA-, AB and AC-stacking, respectively. In addition, the calculated relative energy and layer distance with optB88 vdW functional are consistent with the DFT-D2 ones, this further demonstrates the validity of our calculation. Due to the small energy difference, these configurations are expected to transform to each other under appropriate condition. Therefore, it is desired to investigate electronic structure of different stacking order. As listed in Table 3, the most stable configuration AD-stacking and metastable AA-stacking are metallic, however, the AB- and AC-stacking are predicted to be semiconductors with bandgap of 0.22 and 0.01 eV, respectively. For trilayer $C_3N$, we consider four kinds of high symmetry stacking structures based on the three stacking orders of bilayer $C_3N$: ADA-stacking, ADB-stacking, ADC-stacking, and ADD-stacking. The results show that ADA-stacking is the most stacking order and corresponding structure shows metallic properties. On the whole, the electronic structure of multi-layered $C_3N$ is closely related to the number of layers.

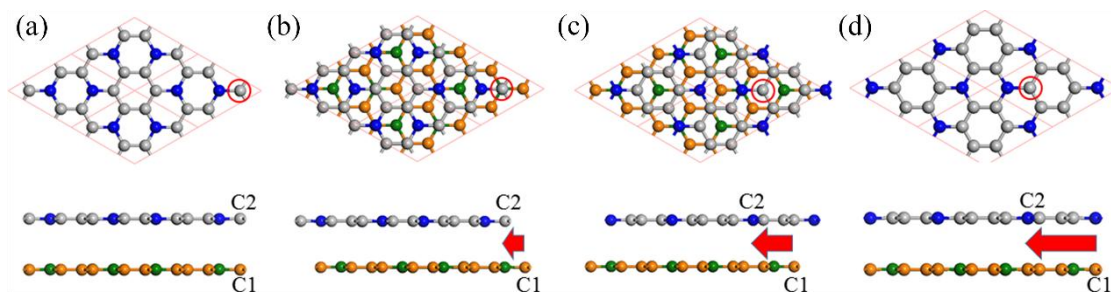

**Fig. 5.** Four stacking structures of bilayer C$_3$N, (a)-(d) Top view and side view of AA-, AB-, AC- and AD-stacking, respectively. A 2×2 super cell is adopted for the top view. The red arrow indicates the relative shift of C1 and C2 atom. The C2 atom in the top layer is labelled by red circle.

**Table 3.** The relative energy, layer distance and bandgap of bilayer C$_3$N under different functional.

|    | relative energy (eV/unit cell) | | layer distance (Å) | | PBE bandgap (eV) |
| --- | --- | --- | --- | --- | --- |
|    | DFT-D2 | optB88 | DFT-D2 | optB88 |   |
| AA | 0.093 | 0.086 | 3.40 | 3.41 | 0.00 |
| AB | 0.003 | 0.007 | 3.22 | 3.25 | 0.22 |
| AC | 0.017 | 0.017 | 3.20 | 3.20 | 0.01 |
| AD | 0.000 | 0.000 | 3.23 | 3.21 | 0.00 |

## Conclusion

In summary, we conduct a first-principles simulation to investigate the thermal and dynamic stability, mechanical and electronic properties of recently synthesized C$_3$N nanosheet. We have identified C$_3$N monolayer is a semiconductor with suitable bandgap and ultra-high mechanical strength and thermal stability. The calculation also indicates that C$_3$N possesses a fascinating bending Poisson's effect, resulting from the change of hybridization state of N atoms in C$_3$N nanotube. Most of the corresponding nanotubes also present high Young's modulus and semiconducting properties. Finally, multi-layered C$_3$N is predicted to possess metallic properties with AD-stacking. Considering these high stability, superior mechanical strength and semiconducting properties of C$_3$N and related derivative, C$_3$N nanosheet is expected to have promising potentials to compete not only with graphene but also with other 2D materials for various applications, particularly in nanotransistors, fabrication of polymer nanocomposites with superior mechanical response. We hope our research can stimulate future experiments on this subject.

## Author information

### Corresponding Author

jlyang@ustc.edu.cn

**Notes**

The authors declare no competing financial interests.

## Acknowledgement

This paper is financially supported by the National Key Research & Development Program of China (Grant No. 2016YFA0200604), by the National Natural Science Foundation of China (NSFC) (Grants No. 21421063, No. 21233007, and No. 21603205), by the Chinese Academy of Sciences (CAS) (Grant No. XDB01020300) and by the Fundamental Research Funds for the Central Universities (Grant No. WK2060030023). We used computational resources of Super-computing Center of University of Science and Technology of China, Supercomputing Center of Chinese Academy of Sciences, Tianjin and Shanghai Supercomputer Centers.

Supporting Information

# C$_3$N: a Two Dimensional Semiconductor Material with High stiffness, Superior Stability and Bending Poisson's Effect

Haidi Wang,[a] Hong Wu,[a] and Jinlong Yang[*ab]

[a]Hefei National Laboratory of Physical Science at the Microscale, University of Science and Technology of China, Hefei, Anhui 230026, China

[b]Synergetic Innovation Center of Quantum Information & Quantum Physics, University of Science and Technology of China, Hefei, Anhui 230026, China

Email: jlyang@ustc.edu.cn

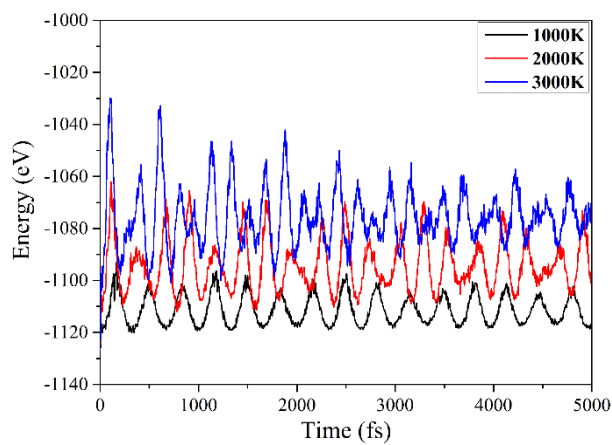

Fig. S1. AIMD simulation of C$_3$N under 1000K, 2000K and 3000K

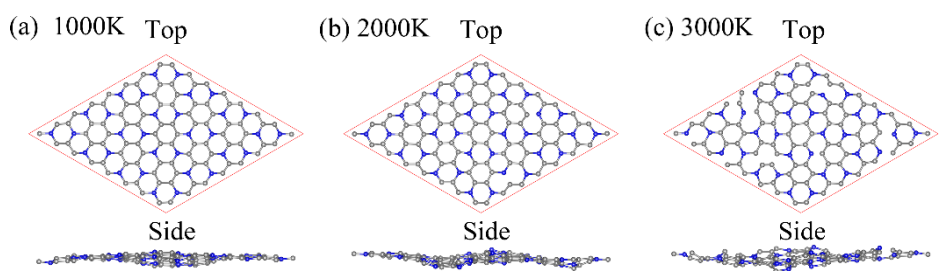

Fig. S2. (a), (b) and (c) the snapshot of C$_3$N's atomic configuration at the end AIMD simulation under 1000K, 2000K and 3000K, respectively.

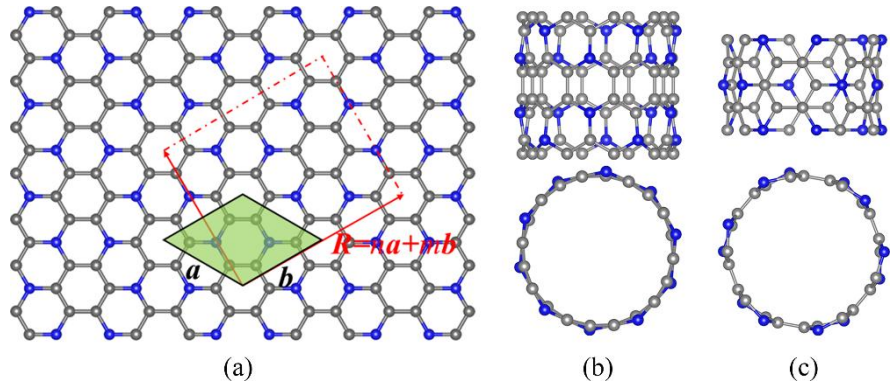

Fig. S3. Schematic plots of C₃N-nanotube which can be viewed as rolling up a C₃N sheet following the roll-up vector **R**=*na*+*mb*. Top and side view of (b) zigzag (5,0) and (c) armchair (-3,3) tube.

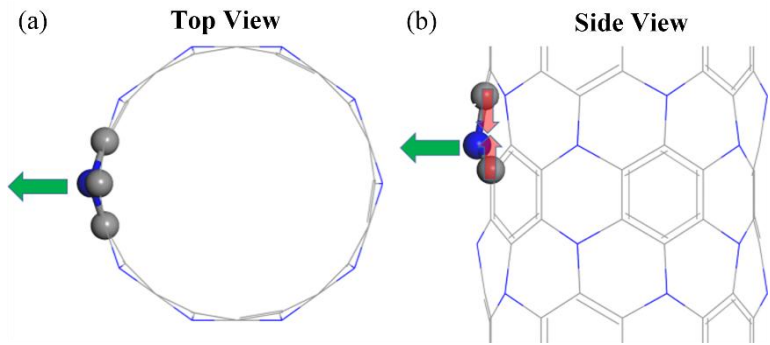

Fig. S4. Local structure of C₃N nanotube. The green arrow indicates the movement of N atom and red arrow presents the contraction of C-C distance.

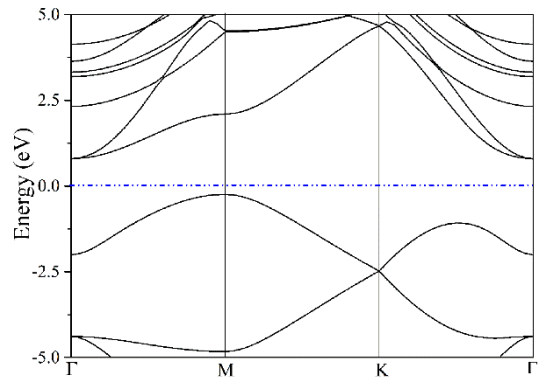

Fig. S5. Electronic band structure of C₃N with HSE06 level of theory.

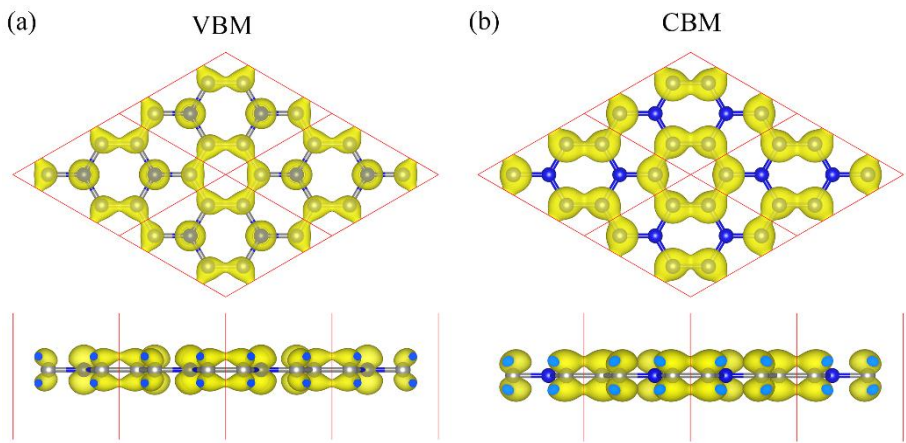

Fig. S6. Electronic density distribution of (a) the valence band maximum and (b) the conduction minimum.

$d_1$, $d_2$, $d_3$, $Y_{zz}$ and $v_{zz}$ are elastic constant related variables.

$$\begin{cases} v_{zz} = \dfrac{C_{12}}{C_{22}} \\ d_1 = \dfrac{C_{11}}{C_{22}} + 1 - \dfrac{C_{11}C_{22} - C_{12}^2}{C_{22}C_{66}} \\ d_2 = -(2\dfrac{C_{12}}{C_{22}} - \dfrac{C_{11}C_{22} - C_{12}^2}{C_{22}C_{66}}) \\ d_3 = \dfrac{C_{11}}{C_{22}} \\ Y_{zz} = \dfrac{C_{11}C_{22} - C_{12}^2}{C_{22}} \end{cases}$$